\documentclass[11pt,english]{article}
\usepackage{amssym}
\usepackage{graphics}
\usepackage{babel}
\usepackage[latin1]{inputenc}
\renewcommand{\baselinestretch}{2.00}
\newcommand{\vs}{\vspace*{0.5in}}

\begin{document}
\vs
\begin{center}
{\Large\bf The impact and rotational lightcurves of Comet 9P/Tempel 1}\\
\vs
{\bf Jean Manfroid}$^{1}$, {\bf Damien Hutsem\'ekers}$^{1}$, {\bf Emmanu\"el Jehin}$^{2}$, \\
{\bf Anita L. Cochran}$^{3}$, {\bf Claude Arpigny}$^{1}$, {\bf William M. Jackson}$^{4}$, \\
{\bf Karen J. Meech}$^{5}$,{\bf  Rita Schulz}$^{6}$, {\bf Jean-Marc~Zucconi}$^{7}$\\
\end{center}
\vs
1. Institut d'Astrophysique et de G\'eophysique, Sart-Tilman, B-4000 Li\`ege, Belgium; manfroid@astro.ulg.ac.be; arpigny@astro.ulg.ac.be; hutsemekers@astro.ulg.ac.be\\
2. European Southern Observatory, Casilla 19001, Santiago, 
Chile; ejehin@eso.org\\
3. Department of Astronomy and McDonald Observatory, University 
of Texas at Austin, C-1400, Austin, USA; anita@barolo.as.utexas.edu\\
4. Department of Chemistry, University of California, 1 Shields Avenue, Davis, CA 95616; wmjackson@ucdavis.edu\\
5. Institute for Astronomy, University of Hawaii at Manoa, 2680 Woodlawn Drive, Honolulu, USA; meech@IfA.Hawaii.Edu \\
6. ESA/RSSD, ESTEC, P.O. Box 299, NL-2200 AG Noordwijk, 
The Netherlands; rschulz@rssd.esa.int\\
7. Observatoire de Besan\c{c}on, F25010 Besan\c{c}on Cedex, 
France; jmz@dalai-zebu.org\\
\vs

\clearpage

Proposed Running Head : The impact and rotational lightcurves of Comet 9P/Tempel 1 \\

First Author address :\\

Jean Manfroid\\
Institut d'Astrophysique et de G\'eophysique\\ 
All\'ee du six ao\^ut 17, Sart-Tilman\\ 
B-4000 Li\`ege, Belgium\\
manfroid@astro.ulg.ac.be

\clearpage

\section*{Abstract}
 
UVES and HIRES high-resolution spectra of comet 9P/Tempel 1 are used
to investigate the impact and rotational light curves of various
species with a view toward building a simple model of the distribution
and activity of the sources.  The emission by  OH, NH, CN, C$_3$, CH,
C$_2$, NH$_2$ and O\,I, are analyzed, as well as the light scattered
by the dust. It is found that a simple model reproduces fairly well
the impact lightcurves of all species combining the production of the 
observed molecules and the expansion of the material
throughout the slit.  The impact light curves are consistent with
velocities of 400--600 m/s. Their modelling requires a three-step
dissociation sequence ``Grand-Parent $\rightarrow$ Parent
$\rightarrow$ Daughter'' to produce the observed molecules.  The
rotational light curve for each species is explained in terms of a
single model with three sources. The dust component can however not easily 
be explained that way.

\vspace{0.7in}

{\bf Keywords}: 
Deep Impact $-$ Comets $-$ 9P/Tempel 1

\clearpage

\section{Introduction}
\label{sec:intro}

The Deep Impact (DI) experiment was a first in cometary
astrophysics.  Instead of observing objects in a relatively steady
state, it was possible to watch the instantaneous release of a large
quantity of cometary material and its subsequent evolution in timescales
of hours and days (A'Hearn et al. 2005).  The fact that the time of origin
of the observed effects is known, provided an extraordininary advantage
in studying the complex evolutive mechanisms governing the physics and 
chemistry in the coma.
The international collaboration also provided the unique opportunity
to monitor a comet during weeks with an armada of telescopes (Meech et
al. 2005).

Here, we present the analysis of data obtained over 15 nights
spead before, during and after the Deep Impact event (2005 July 4, 
05:52 UT as seen from Earth) with the high-resolution spectrographs of
the ESO VLT (European Southern Observatory, Very Large Telescope,
Cerro Paranal, Chile) and Keck (Hawaii) observatories.  We investigate
the impact and the rotational light curves of various emissions and
provide a simple model for the activity of the sources.  The
emission by   OH, NH, CN, C$_3$, CH, C$_2$, NH$_2$ and O\,I, are
analyzed, as well as the light scattered by the dust.
 
The observations are presented in Section \ref{sec:obs}. The light
curves observed immediately after the impact at the Keck I telescope
are presented in Section \ref{sec:impact} together with their 
interpretation by a simple model. In Section \ref{sec:rota}, these light 
curves are again used to build a model also for the rotational light 
curves that resulted from the data collected during 13 nights at the VLT. 
Finally we summarize our conclusions in Section \ref{sec:conclu}.

\section{Observations and data reduction}
\label{sec:obs}

The high-resolution spectra of comet 9P/Tempel 1 obtained with the VLT
UVES (ESO UltraViolet Echelle Spectrograph) and Keck I HIRES (High
Resolution Echelle Spectrograph) between May and July 2005 are
described in Jehin et al (2006), and in Tables~\ref{tab:circonsa} and
\ref{tab:circonsb}.  The slits used at both telescopes had similar areas 
($10''$ by $0.44''$ for UVES, $7''$ by $0.86''$ for HIRES) and were 
generally centered on the nucleus, which allows some comparison of the 
photometric light curves.

The reductions were made in the usual manner for cometary spectra.
The echelle package of IRAF (Image Reduction and Analysis Facility)
and the UVES pipeline (v1.4) implemented 
in the ESO-MIDAS (Münich Image Data Analysis System)
software package (Ballester et al. 2000) were used to extract 
the various orders. The dust and twilight contributions were subtracted 
after proper scaling and Doppler shifting of an appropriate solar (UVES) 
or solar analog (HIRES) spectrum. This was done order by order so that linear 
wavelength shifts and simple scaling factors could be used.
The latter was derived in an iterative procedure until nulling of all 
remaining solar features.

The spectrographs were not designed to be used as high precision
photometers, and the observations were not conducted with the
photometric issue as a priority. The centering on the nucleus was not
always perfect and the slit width was only a fraction of an arc
second. Some UVES spectra were intentionally offset from the nucleus
by between 1 and 2 arcsec (see Table~\ref{tab:circonsa}).  This can
lead to significant effects on components with a sharp spatial profile
(dust). The position of the nucleus along the slit was not quite
stable in the HIRES data. A slight drift along the slit occured during
the impact night, partly because of the differential atmospheric
refraction (the slit was aligned along the parallactic angle and there
was no refraction corrector).  At high air mass, in the bluest orders
of the last spectra, the nucleus was almost at the edge of the slit.

The main problem was thus to choose the proper extraction aperture.
Calculations made with different apertures showed differences of up to
15 \% in the photometric light curve during the impact night.
After some testing we found that the most sensible solution was to use the 
full slit length on both spectrographs, keeping in mind the intrinsic limitations 
of such measurements. This full aperture could be easily defined from the 
flat frames and it yielded the maximum signal-to-noise ratio, which is 
certainly a determining factor for such a faint comet.

\begin{figure}
\includegraphics*{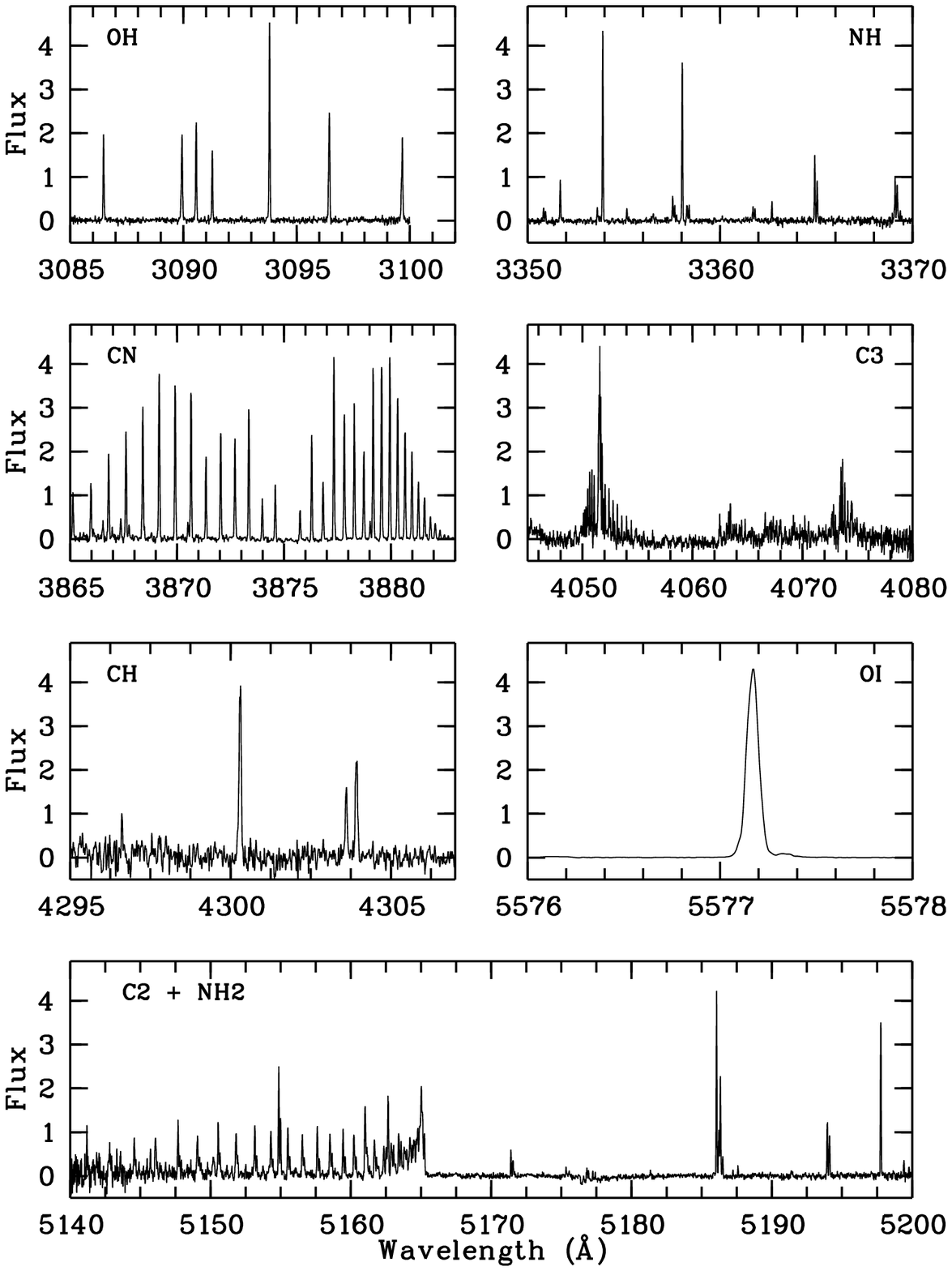}
      \caption{Spectral features used to measure the various species. 
They are extracted from spectra taken with UVES the night 
following the impact. The flux is given in arbitrary units.
The cometary [OI] is the small bump on  the red wing
of the strong telluric line. Most of the band blueward of 517~nm
is C$2$, while most of the lines above 518~nm are NH$_2$ ([NI] 519.8~nm
is telluric). 
 }
\label{figspectres}
   \end{figure}

From these spectra we derived the flux of major emission bands and of the
dust scattered solar spectrum, assuming a clear atmosphere, and stable
detectors. Eight gaseous species were measured and analyzed : OH (0-0) at 
308 nm, NH (0-0) at 336 nm, CN (0-0) at 388 nm, C$_3$ complex at 405 and 
407 nm, CH (0-0) at 430 nm, C$_2$ (0-0) band head at 516 nm, various lines
of NH$_2$ at 519, 531, 569, 574 nm, and the green and red oxygen forbidden 
lines ($[$O\,I$]$) at 558 nm and 630 nm (UVES only). Table~\ref{tab:gasspec}
details the 
wavelength ranges of interest where major emission features
and the corresponding comparison zones (not always emission free) can be found. 
The wavelengths are given in the rest frame of the nucleus.
For OH, NH and CN the integration
was done on  a series of individual lines. The [O\,I] lines have been deblended 
from the telluric features.  
Plots of a typical spectrum are shown on 
Fig.~\ref{figspectres}.\\

UVES data were obtained over 13 nights (10 consecutive nights encircling 
the impact date). This provided complete coverage of the comet's light curve
over several rotational periods.  As the CN band was covered in
all UVES setups, it is the species with the most observations. Other
molecules have less dense coverage in the phase diagram.  After
correcting for the higher efficiency of new mirror coatings 
and for the strong Swings effect closer to the comet's perihelion, 
we could directly compare CN fluxes measured in June and July 2005. The
comet activity appeared to be slightly higher (about 3 \%) in June, for which
we corrected the CN data. However, for the other species we used only the
July observations, as we could not do reliable adjustments owing to less 
available data and/or the lack of accurate models.

The dust contribution has been accurately evaluated in each spectrum 
in order to perform an accurate continuum subtraction. However, as the small
slit width of UVES lead to large variations when the slit was not perfectly 
centered on the nucleus and/or when the seeing changed, we kept only the well-centered
spectra (see Table~\ref{tab:circonsa}) for evaluation. The HIRES data were 
obtained with a wider slit and therefore did not show this effect.
The main HIRES data set covers the four hours following the impact (see Table 3).
In addition three HIRES spectra obtained on May 30 are available
\footnote{Those data are publicly available at http://msc.caltech.edu/deepimpact/}.

The relative variations in the brightness of the dust appear to be wavelength independent
over the whole spectral range, both in the UVES and HIRES impact 
data. The data presented here are the same as in Jehin et al (2006).

\section{Modelling the impact light curve}
\label{sec:impact}

\begin{figure}
\hspace{-15mm}\includegraphics*{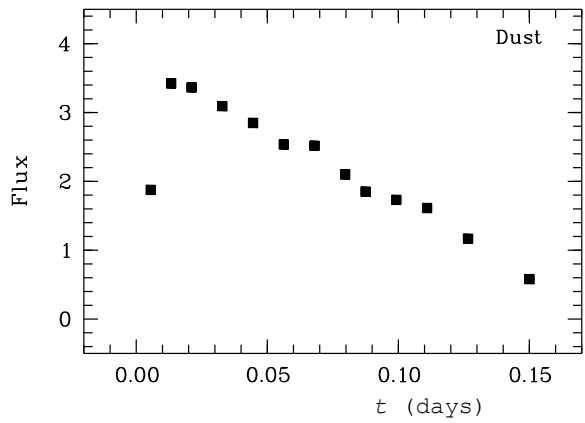}
\caption{The observed dust light curve in arbitrary units.}
\label{fig:i_dust}
\end{figure}

The impact light curves were deduced from the HIRES and UVES spectra
obtained during the hours (or nights) following the impact.
They had to be corrected for the natural rotational variations, which
were most important for the spectra obtained during the
first hours after the impact as they coincide with the very phase
when the natural light curves exhibit the fastest variations of the
1.7 d cycle.  The resulting impact light curves are illustrated in
Figs.~\ref{fig:i_cn}--\ref{fig:i_dust}.

\subsection{A simple model}

We assume that the observed species --the daughter-- originates from a
single parent which itself comes from a single grandparent in a
continouous sequence of dissociation (the three species coexist at any $t$).  
The grandparent appears necessary to
introduce a delay in the light curves (cf. Sect.~3.2).  Within this
framework, we may write
\begin{eqnarray}
\nonumber & & dn_{g} = -\alpha_{g} n_{g} dt \\
& & dn_{p} = +\alpha_{gp} n_{g} dt - \alpha_{p} n_{p} dt  \\
\nonumber & & dn_{d} = +\alpha_{pd} n_{p} dt - \alpha_{d} n_{d} dt 
\end{eqnarray}
from which we derive
\begin{eqnarray}
\frac{n_{d}}{n_{g_{0}}} =
\frac{\alpha_{pd}\,\alpha_{gp}}{(\alpha_{p}-\alpha_{d})(\alpha_{g}-\alpha_{d})} e^{-\alpha_{d} t} +
\frac{\alpha_{pd}\,\alpha_{gp}}{(\alpha_{d}-\alpha_{p})(\alpha_{g}-\alpha_{p})} e^{-\alpha_{p} t} +
\frac{\alpha_{pd}\,\alpha_{gp}}{(\alpha_{d}-\alpha_{g})(\alpha_{p}-\alpha_{g})} e^{-\alpha_{g} t}\,\, .
\end{eqnarray}
In these equations, $n$ is the number of molecules, the subscripts
$d$, $p$, $g$ standing for daughter, parent and grandparent
respectively.  $n_{g_{0}}$ is the number of grandparents produced at
once by the impact. The grandparent can be anything, e.g. icy grains,
complex or simple molecules, as long as it obeys Eq.~1.  $\alpha =
1/\tau$ are inverse lifetimes; $\alpha_{g}$, $\alpha_{p}$ and
$\alpha_{d}$ characterize the destruction or dissociation of the
grandparent, parent and daughter; $\alpha_{gp}$ and
$\alpha_{pd}$ refer to the production of the parent from
the dissociation of the grandparent and to the production of the
daughter from the dissociation of the parent, accounting for the
possibility that the grandparent (resp. parent) may dissociate into
several parents (resp. daughters). We then have $\alpha_{g} = \sum_{p}
\alpha_{gp}$ and $\alpha_{p} = \sum_{d} \alpha_{pd}$.  If the
dissociation lifetime of the daughter is very long compared to the
timescale of the event ($\alpha_{d} \simeq$ 0) and if the grandparent
does not exist or has a very small lifetime, Eq.~2 will simplifly to
\begin{eqnarray}
\frac{n_{d}}{n_{p_{0}}} = 1 - e^{-\alpha_{p} t}
\end{eqnarray}
which is useful to get a first insight into the observed light curves.
In this case, $n_{p_{0}}$ is the number of parents produced by the
impact.

Eqs.~2 and 3 are only valid if all the expanding material is within
the slit area.  When a part of the material moving at velocity $v$
exits the slit width $w$, i.e., at $t > t_w= w/2v$, the number of
particles within the slit decreases with time.  Assuming that the
expanding material forms a shell of radius $v \, t$, the volume inside
and outside the slit area may be readily computed such that for
$t > t_w$
\begin{eqnarray}
n_{d}^{'}  \; = \; n_{d}  \; \frac{t_w}{t}
\end{eqnarray}
where  $n_{d}$ is given by Eq.~2 or 3 (when $t \leq t_w$, $n_{d}^{'} = n_{d}$).
Although established for a spherical expansion, this relation is also
valid for instance for a hemispherical flow.  When $t > t_l$, the material
fills both the slit width $w$ and the slit length $l$.
Assuming  $l\gg w$, we may write
\begin{eqnarray}
n_{d}^{'} \;  = \; n_{d}  \; \frac{t_w}{t} \; 
\frac{2}{\pi} \arcsin (\frac{t_l}{t}) 
\end{eqnarray}
which becomes
\begin{eqnarray}
n_{d}^{'} \;  = \; n_{d}  \; 
\frac{2}{\pi} \; \frac{t_w t_l}{t^2} 
\end{eqnarray}
when $t \gg t_l$. In the case of the HIRES data, $t_l = 8.1 \, t_w$.
Assuming the medium to be optically
thin, the observed flux is simply proportional to $n_d^{'} (t)$
integrated over the exposure time.

Since it is unlikely that all the material is moving at a single
velocity $v$, we introduce a velocity distribution.  For convenience
we use
\begin{eqnarray}
d\,n_{g_{0}} \; =  \; N_{g_{0}} \, \frac{1}{\sigma_{v} \sqrt{\pi}} \;
e^{-(v-\bar{v})^2 / \sigma_{v}^{2}} \,dv
\end{eqnarray}
where the mean velocity $\bar{v}$ and its standard deviation 
$\sigma_{v}$ are parameters which can be estimated
from the observations.  It is important to note that in this
simplified model we assume that the daughter, parent and grandparent
species expand with the same velocity.

To summarize, we have
\begin{eqnarray}
n_{d}^{'} (t) \; = \; \int N_{g_{0}} B \,D(v) \, R(t) \, S(v,t) \, dv 
\end{eqnarray}
integrated over the velocity distribution, with
\begin{eqnarray}
D(v) \; = \; \frac{1}{\sigma_{v} \sqrt{\pi}} \; 
e^{-(v-\bar{v})^2 / \sigma_{v}^{2}} \; ,
\end{eqnarray}
\begin{eqnarray}
R(t) \; = \; 
\frac{\alpha_{p}\,\alpha_{g}}{(\alpha_{p}-\alpha_{d})(\alpha_{g}-\alpha_{d})} e^{-\alpha_{d} t} +
\frac{\alpha_{p}\,\alpha_{g}}{(\alpha_{d}-\alpha_{p})(\alpha_{g}-\alpha_{p})} e^{-\alpha_{p} t} +
\frac{\alpha_{p}\,\alpha_{g}}{(\alpha_{d}-\alpha_{g})(\alpha_{p}-\alpha_{g})} e^{-\alpha_{g} t}
\end{eqnarray}
and\\[0.2cm] \hspace*{3cm}
\begin{eqnarray}
\nonumber & & S(v,t)\; = \; 1  
               \hspace*{3.35cm}    {\rm when\ } t_w \geq t \; \; \\
          & & S(v,t)\; = \;\frac{t_w}{t}  
               \hspace*{3.1cm}     {\rm when\ } t_w <  t {\rm \ and\ } t_l \geq t \; \; \\
\nonumber & & S(v,t)\; = \;\frac{t_w}{t} \; \frac{2}{\pi} \arcsin (\frac{t_l}{t}) 
               \hspace*{1cm}      {\rm when\ } t_l <  t \; \;
\end{eqnarray}
where $t_w = w / 2 v$, $t_l = l / 2 v$ and $B = \alpha_{pd}
\alpha_{gp} / \alpha_{p} \alpha_{g}$.  Free parameters are $\bar{v}$,
$\sigma_{v}$, $\alpha_{d}$, $\alpha_{p}$, $\alpha_{g}$ and $N_{g_{0}}
B$.  Since absolute fluxes are not measured, $N_{g_{0}} B$ is just a
scaling factor.  As seen in Eq.~10, $R(t)$ is identical if we
exchange the values of $\alpha_{p}$ and $\alpha_{g}$, such that
parameter fitting can be done with either $\alpha_{g} / \alpha_{p} >$
1 or $\alpha_{g} / \alpha_{p} <$ 1.

\subsection{Interpretation of the light curves}

\begin{figure}[t]
\hspace{-15mm}\includegraphics*{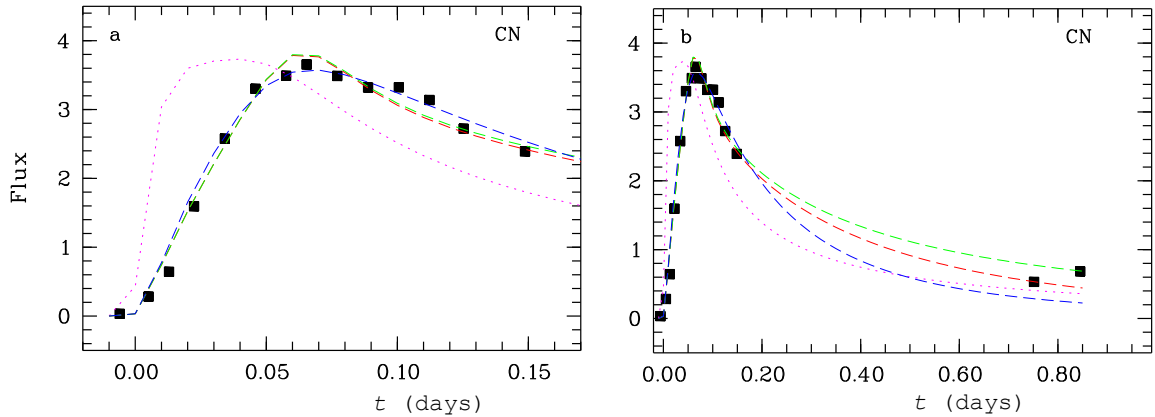}
\caption{The observed CN light curve in arbitrary units.  
In this figure and in the following ones, $t$ is the time 
elapsed since the impact.  The
superimposed dashed lines are examples of model fitting. In all cases
$\bar{v}$ = 400 m/s and $\alpha_d$ = 0.2 d$^{-1}$.  Model 1 (red) is
computed with $\sigma_v$ = 100 m/s, $\alpha_p$ = 2 d$^{-1}$ and
$\alpha_g / \alpha_p$ = 4.  Model 2 (green) is computed with
$\sigma_v$ = 100 m/s, $\alpha_p$ = 0.1 d$^{-1}$ and $\alpha_g /
\alpha_p$ = 100.  Model 3 (blue) is computed with $\sigma_v$ = 200
m/s, $\alpha_p$ = 5 d$^{-1}$ and $\alpha_g / \alpha_p$ = 4.  Model 4
(dotted pink) has no grandparent ($\sigma_v$ = 200 m/s, $\alpha_p$ =
0.1 d$^{-1}$) and is clearly unable to reproduce the data.
Fig.~\ref{fig:i_cn}b shows the better agreement of model 1 with the
two later UVES measurements (scaled to the HIRES slit area).}
\label{fig:i_cn}
\end{figure}

{\bf CN --} Fig.~\ref{fig:i_cn} illustrates the observed post-impact
light curve of CN (hereafter the daughter molecule).  It is
characterized by a rather steep rise and a smooth decline.  The
maximum corresponds to the time at which the material fills either the
slit width or the slit length. At times $t > t_w$ and $t > t_l$, the
light curve can be seen as a combination of the production of CN
(Eq.~2) and the dilution of the expanding material beyond the slit
area (Eqs.~4 or 5).  At large $t$ we may neglect, to first
approximation, the effect of the grandparent as well as the CN
dissociation lifetime, which is very long 
(Fray et al. 2005 and references therein) --- probably 
longer than 4 days (Keller et
al. 2005). Therefore Eq.~3 holds at large $t$. In this case, the number 
of daughter molecules may be constant after some time 
(as long as $\alpha_p\,t\ll 1$ in Eq. 3)
or proportional to $t$ (if $\alpha_p\,t\gg 1$). We then expect the shape
of the decline of the light curve to be roughly intermediate between a
constant and $t^{-1}$ if the bump corresponds to the filling of the
slit width (Eqs. 3 \& 4) or between $t^{-1}$ and $t^{-2}$ if the
bump corresponds to the filling of the slit length (Eqs. 3 \&
6). However, since the velocity of the flow is not unique the
situation is less cut-clear and an independent knowledge of the bulk
velocity of the flow is needed.

In the case of CN, the analysis of the spatial profiles has led us
(Jehin et al. 2006) to estimate the
velocity of the expanding material to $\bar{v} \simeq$ 400 m/s
(with a range of projected velocites from 250 to 650 m/s).  Since
the $0.86''\times7.0''$ HIRES slit corresponds to 562 km $\times$ 4570
km at the comet at the epoch of the observations, the slit length is
filled after $t_l = 0.066$~d, which corresponds to the bump seen in the
light curve. The fact that the decline of the light curve is not as
steep as $t^{-2}$ indicates that the CN production still continues at $t >
t_l$ which requires a rather long lifetime for its parent.  If the
material is expanding at a single velocity ($\sigma_v \ll \bar{v}$), we will
expect a sharp decline just after filling the slit, i.e., a relatively
narrow bump. The fact that such a bump is not observed requires some
smoothing, i.e. a large velocity dispersion. After some trial we find
that $\sigma_v \gtrsim$ 100 m/s is needed to reproduce the data, which is
in agreement with initial  estimates (Jehin et al. 2006).

As illustrated in Fig.~\ref{fig:i_cn}, modelling the CN light curve
with only a parent molecule results in a much faster rise after the
impact and a clear discontinuity at $t \simeq t_w$, when the
material leaves the slit width. Such models cannot reproduce the data
whatever values are adopted for the lifetimes. A grandparent must
be added to introduce a delay in the light curve just after the
impact. In Fig.~\ref{fig:i_cn}a, we show reasonable fits to the data
with different combinations of $\alpha_p$ and $\alpha_g / \alpha_p$,
indicating some degeneracy between $\alpha_p$ and $\alpha_g$. The
value of the parent lifetime needed to reproduce the data, does not
only depend on the ratio $\alpha_g / \alpha_p$, but also on the adopted
velocity distribution. Fig.~\ref{fig:i_cn}a shows an equally
acceptable fit with $\sigma_v$ = 200 m/s. In general, adding material
at lower velocities enhances the number of molecules in the slit at $t
> t_l$, and consequently requires a shorter parent lifetime to fit the data.

Although we are able to reproduce fairly well the observed light curve,
the parent lifetime $\tau_p = 1 / \alpha_p$ that we derive appears
very uncertain owing to the strong dependence on both the velocity
distribution and the ratio $\alpha_g / \alpha_p$. Without additional
data and/or independent constraints, the CN parent dissociation
lifetimes, whcih we estimate from acceptable fits, are roughly in the range
0.1 -- 10 d.  Fortunately, additional measurements from UVES
observations were performed several hours after the impact allowing us
to constrain further the model (Fig.~\ref{fig:i_cn}b). We find that
models with 0.2 d$^{-1}$ $\lesssim \alpha_p \lesssim$ 4 d$^{-1}$ are
favoured, which corresponds to CN parent dissociation lifetimes of 2
10$^{4}$ s $\lesssim \tau_p \lesssim$ 5 10$^{5}$~s. Even with this
additional constraint, the derived lifetimes are accurate only within
orders of magnitude. They are consistent with other estimates (Keller et
al. 2005, Fray et al. 2005) and with the hypothesis that 
HCN is the parent of CN.  The
CN grandparent destruction lifetime is even more uncertain, providing
no information on its nature.\\

\begin{figure}
\hspace{-15mm}\includegraphics*{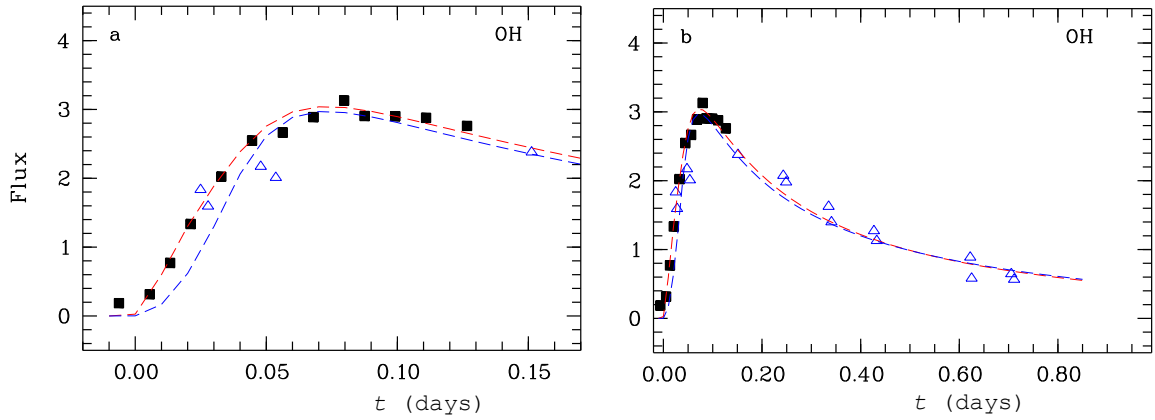}
\caption{The observed OH light curve in arbitrary units. The triangles
are scaled data from Schulz et al. (2006).  
Model 1 (red) is computed with $\bar{v}$ =
400 m/s, $\sigma_v$ = 200 m/s, $\alpha_d$ = 0.4 d$^{-1}$, $\alpha_p$ =
0.5 d$^{-1}$ and $\alpha_g / \alpha_p$ = 40. Model OM1 (blue) uses the
same parameters but it is adapted to the circular aperture of 6''
(3900 km) used by Schulz et al. (2006) and scaled to correspond to
model 1 at large $t$ (with a circular aperture, Eq.~5 must be
replaced by $n_{d}^{'} \; = \; n_{d} \; (1-\sqrt{1-(t_w/t)^2})$).}
\label{fig:i_oh}
\end{figure}

\begin{figure}
\hspace{-15mm}\includegraphics*{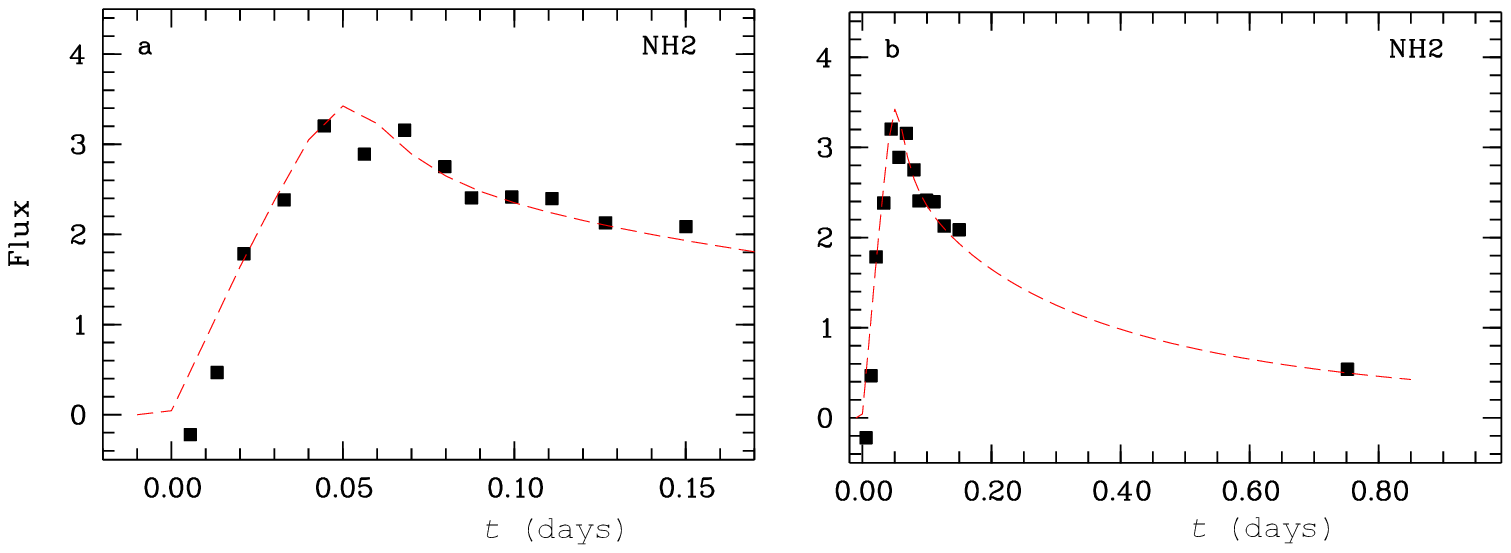}
\caption{The observed NH$_2$ light curve in arbitrary units.
The illustrated model
 is computed with $\bar{v}$ = 500 m/s, $\sigma_v$ = 100 m/s,
$\alpha_d$ = 0.6~d$^{-1}$, $\alpha_p$ = 10 d$^{-1}$ and $\alpha_g /
\alpha_p$ = 0.05. }
\label{fig:i_nh2}
\end{figure}

\begin{figure}
\hspace{-15mm}\includegraphics*{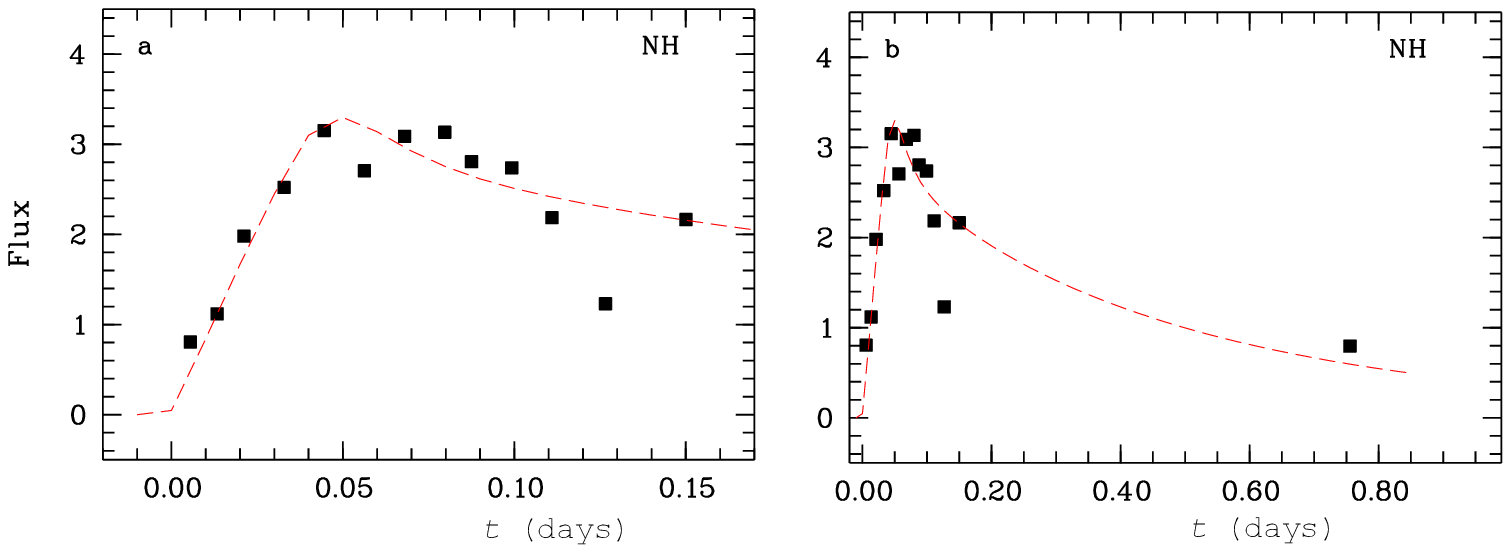}
\caption{The observed NH light curve in arbitrary units.
The illustrated model
 is computed with $\bar{v}$ = 550 m/s, $\sigma_v$ = 150 m/s,
$\alpha_d$ = 0.8~d$^{-1}$, $\alpha_p$ = 2 d$^{-1}$ and $\alpha_g /
\alpha_p$ = 2. }
\label{fig:i_nh}
\end{figure}

\begin{figure}
\hspace{-15mm}\includegraphics*{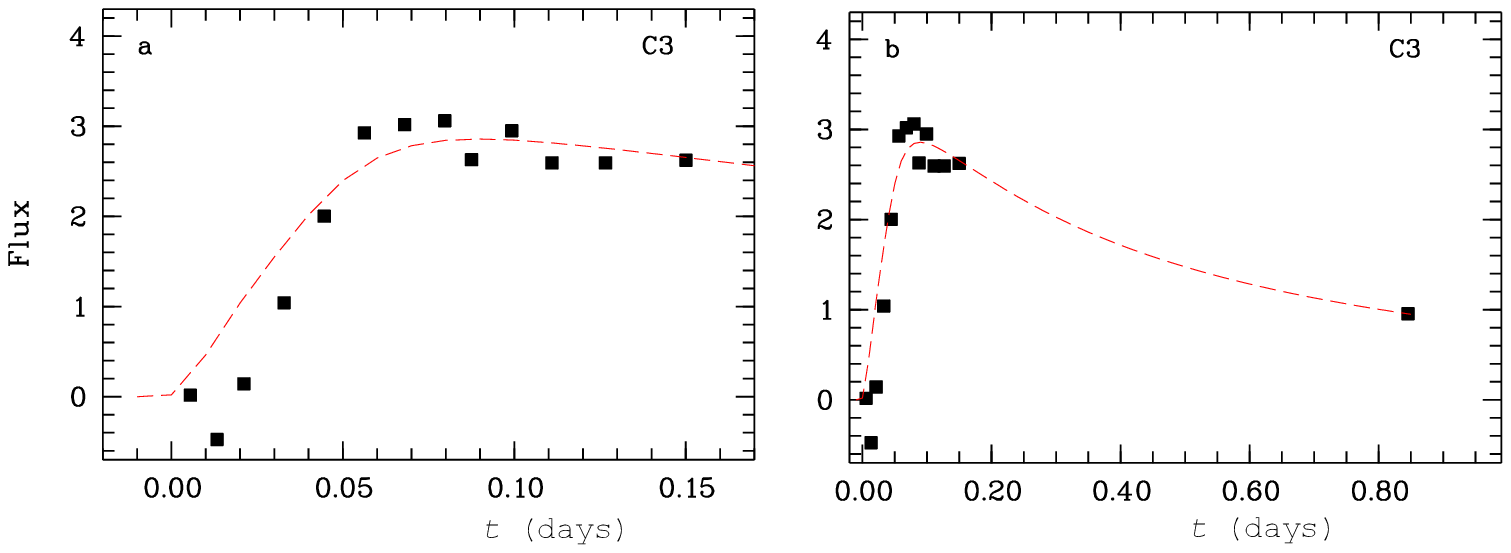}
\caption{The observed C$_3$ light curve in arbitrary units.
The illustrated model
is computed with $\bar{v}$ = 400 m/s, $\sigma_v$ = 200 m/s,
$\alpha_d$ = 0.3~d$^{-1}$, $\alpha_p$ = 10 d$^{-1}$ and $\alpha_g /
\alpha_p$ = 0.01.}
\label{fig:i_c3}
\end{figure}

\begin{figure}
\hspace{-15mm}\includegraphics*{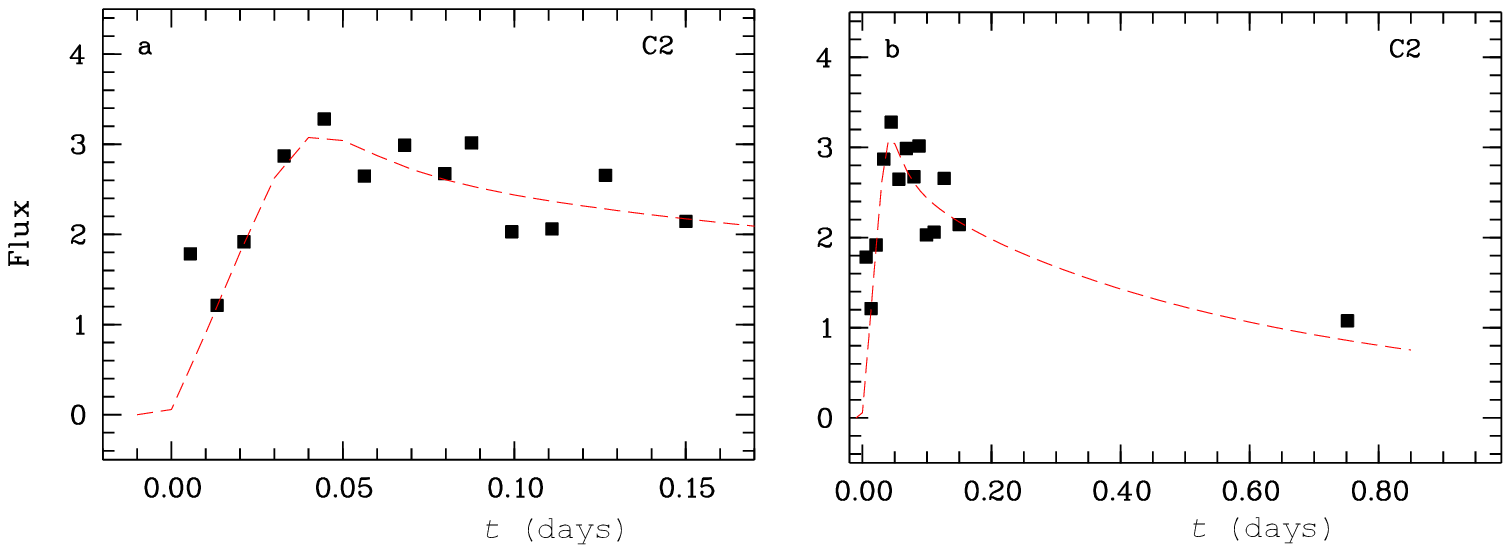}
\caption{The observed C$_2$ light curve in arbitrary units.
The illustrated model
is computed with $\bar{v}$ = 650 m/s, $\sigma_v$ = 200 m/s,
$\alpha_d$ = 0.3~d$^{-1}$, $\alpha_p$ = 1 d$^{-1}$ and $\alpha_g /
\alpha_p$ = 4. }
\label{fig:i_c2}
\end{figure}

\begin{figure}
\hspace{-15mm}\includegraphics*{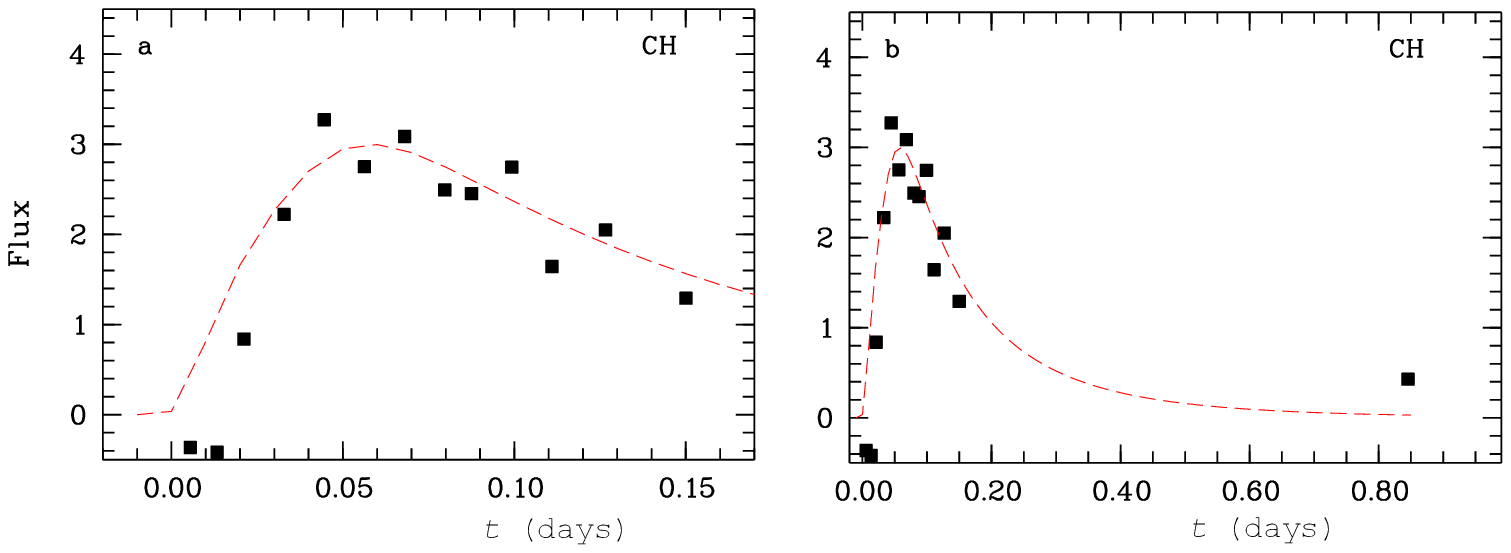}
\caption{The observed CH light curve in arbitrary units.  
The illustrated model is
computed with $\bar{v}$ = 400 m/s, $\sigma_v$ = 200 m/s, $\alpha_d$ =
8.0~d$^{-1}$, $\alpha_p$ = 2 d$^{-1}$ and $\alpha_g / \alpha_p$ =
15. }
\label{fig:i_ch}
\end{figure}

\begin{figure}
\hspace{-15mm}\includegraphics*{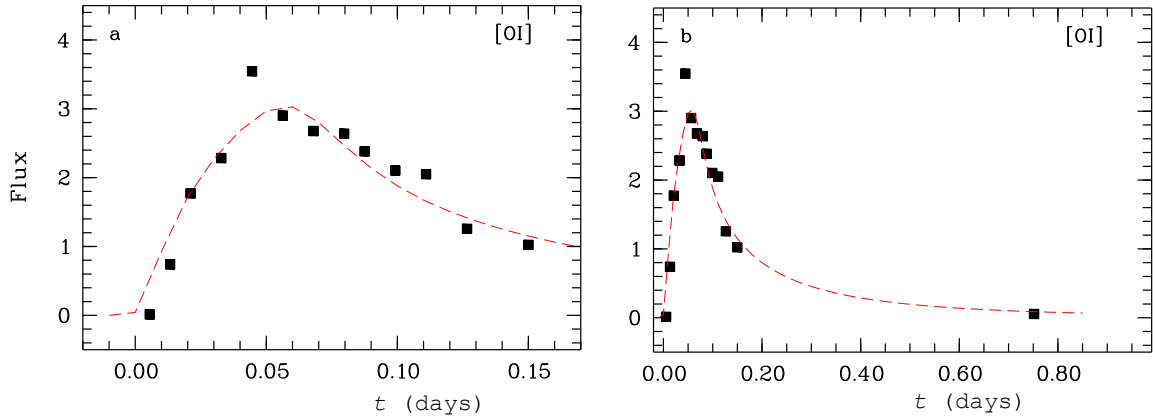}
\caption{The observed [OI] 5577 \AA\ light curve in arbitrary units.
The illustrated model
 is computed with $\bar{v}$ = 400 m/s, $\sigma_v$ = 100 m/s,
$\alpha_d$ = 0.4 d$^{-1}$, $\alpha_p$ = 4 d$^{-1}$ and $\alpha_g / \alpha_p$ =
10. }
\label{fig:i_o1}
\end{figure}

Figures~\ref{fig:i_oh}--\ref{fig:i_o1} illustrate the light curves of
the other molecules and [OI]. The modelling is essentially similar to
the one discussed in detail for CN. 
When necessary, the velocity distribution is fine tuned
in order to reach the best fit. Indeed, differences can arise  
in, e.g., the dissociation process.
Additional details are briefly
described below while the derived parameters are gathered in
Table~\ref{tab:results}.\\


{\bf OH --} In Fig.~\ref{fig:i_oh} the photometry obtained with the
Optical Monitor (OM) of the XMM-Newton satellite (Schulz et al. 2006)
has been included.  The simultaneous fit of HIRES and XMM-OM data
requires a velocity of the OH radicals comparable to the velocity of
CN. We then adopt 400 m/s for  the material to exit the HIRES slit
length at $t_l$ = 0.066 d and the XMM-OM aperture at $t_w$ = 0.056 d.
Models of the light curve are not very sensitive to the long OH
dissociation lifetime ($\tau_d \sim$ 2 10$^{5}$ s at $r_h = 1.5$ AU,
from Morgenthaler et al. 2001) and  require both a grandparent and a
parent (Fig.~\ref{fig:i_oh}).  Acceptable models reproduce fairly well
the data with the parent and grandparent destruction lifetimes given
in Table~\ref{tab:results}, which, as in the case of CN, appear to be
poorly determined.  The OH parent dissociation lifetimes we derive
largely encompass the dissociation lifetime of H$_{2}$O ($\sim$ 8
10$^4$ s), a possible parent of OH.\\


{\bf NH$_2$ --} The relatively narrow bump in the NH$_2$ light curve
(Fig.~\ref{fig:i_nh2}) suggests a slightly higher velocity with a
not too high dispersion. We adopt $\bar{v}$ = 500 m/s and $\sigma_v$ =
100~m/s.  A range of acceptable combinations of $\alpha_p$ and
$\alpha_g$ is given in Table~\ref{tab:results}. The long NH$_2$
dissociation lifetime has little effect on the models ($\tau_d >$ 1.5 10$^{5}$ s at
$r_h = 1.5$ AU, from Cochran et al. 1992). The derived NH$_2$ parent
lifetimes are poorly determined and are consistent with other estimates
(e.g. Wyckoff et al.  1988, Fink et al. 1991).\\


{\bf NH --} A slightly higher velocity is needed to
reproduce the data; we adopt $\bar{v}$ = 550 m/s and $\sigma_v$ =
150~m/s (Fig.~\ref{fig:i_nh}).  The long NH dissociation lifetime has
little effect on the models ($\tau_d >$ 1 10$^{5}$~s from Schleicher
et al. 1989). The derived NH parent dissociation lifetimes are
consistent with other estimates (e.g. Schleicher et al. 1989).\\


{\bf C$_{3}$ --} Keeping in mind the large uncertainties of the flux
measurements (cf. the measurements just after the impact), acceptable
models can reproduce the data with $\bar{v} \simeq$ 400 m/s
(Fig.~\ref{fig:i_c3}).  A range of acceptable combinations of
$\alpha_p$ and $\alpha_g$ is given in Table~\ref{tab:results}. The
long C$_{3}$ dissociation lifetime has little affect on the models
($\tau_d >$ 3 10$^{5}$~s at $r_h = 1.5$ AU from Cochran et
al. 1992). The rather short C$_{3}$ parent dissociation lifetimes we
derive are consistent with other estimates (e.g., Rauer et al. 2003,
Randall et al. 1992).\\


{\bf C$_2$ --} A higher velocity is needed to reproduce the
data; we adopt $\bar{v} = 650$~m/s and $\sigma_v = 200$~m/s
(Fig.~\ref{fig:i_c2}). However the measurements are very uncertain and
should be seen with caution.  The long C$_2$ dissociation lifetime has
little effect on the models ($\tau_d > 3\,10^{5}$~s at $r_h = 1.5$~AU
from Cochran et al. 1992). The derived C$_2$ parent dissociation
lifetimes are consistent with other estimates (Combi and Fink 1997 and
references therein).\\


{\bf CH --} While the range of the adopted CH parent and grandparent
lifetimes (Table~\ref{tab:results}) reproduces reasonably well the
major part of the light curve (given the uncertainties) and agrees
with other estimates (e.g. Cochran et al. 1992), the relatively short
CH dissociation lifetime used in the modelling ($\tau_d >$ 1
10$^{4}$~s at $r_h = 1.5$ AU, from Cochran et al. 1992) might be
underestimated since the residual flux at $t\simeq$ 0.85 d cannot be
fitted by the model (Fig.~\ref{fig:i_ch}b).  However the measurements
for CH are very uncertain and should be regarded with caution.\\


{\bf OI --} Photodissociation of molecules like H$_2$O, OH, CO$_2$ or
CO can produce O\,I directly in the excited $^1$D and $^1$S states
(Festou and Feldman 1981, Morgenthaler et al. 2001). These metastable
states have a lifetime of $\sim110$~s and $\sim0.74$~s, respectively,
much shorter than the photodissociation lifetime of the parent
molecules.  Those atoms travel only a short distance before decaying
to the ground state and emitting photons at 6300\AA\ (and 6364 \AA)
and at 5577\AA\ respectively.

In the modelling procedure, we first considered OI atoms as the
daughter species. However, given the very short lifetime of the
metastable levels, the light curve actually traces the parent molecule,
such that one stage in the 3-step dissociation sequence used up to now
is lost.  Such models provide poor fits essentially because the
computed light curves are rising too fast just after the impact. When
considering either OH or CO as the ``daughter''species --with rather 
long dissocation lifetimes-- and H$_2$O or CO$_2$ as the ``parents'', 
coming themselves from unknow grandparents released by the impact,
acceptable fits of the light curve are obtained (Fig.~\ref{fig:i_o1}) with 
various combinations of $\alpha_p$ and $\alpha_g$ (Table~\ref{tab:results}).
While the information the derived lifetimes provide on the parent 
species is only limited, the fact that a 3-step dissociation sequence is
needed to reproduce the data suggests that OI~($^1$S) atoms are not
produced directly from the dissociation of H$_2$O or CO$_2$, but
through OH or CO.  Moreover, the OI light curve is not similar to the 
one of OH (Fig.~\ref{fig:i_oh}). This might indicate a significant 
contribution from the CO$_2$ $\rightarrow$ CO dissociation sequence.

\subsection{The dust light curve}

The variations of the dust after the impact are much faster than for the
gas, with a
very steep brightening (Fig.~\ref{fig:i_dust}).  The maximum is
already reached at about UT 06:18 $\pm$ 10\,min, the dust being at
that time enhanced by a factor of $\sim8.5$ with respect to the
pre-impact spectrum.  The decline of the dust emission may also be
interpreted as the escape of dust from the slit. This would give a
projected dust velocity of $\sim0.18 \pm$~0.05 km/s, as, in this case,
the escape from the slit width is the dominant factor.  
However
the slow and quasi linear decline would require a rather broad range
of velocities -- slower than the gas component -- and/or complex
processes like destruction of highly reflective icy grains by
sunlight.
Similar dust outflow projected velocities were deduced from the 
analysis of the radial profiles
($\sim0.13 \pm$~0.03 km/s, Jehin et al. 2006) and from the
evolution of the dust cloud ($\sim0.18-0.2$~km/s, Meech et al. 2005 and references
therein; $\sim0.13-0.23$~km/s Schleicher et al. 2006), .

\section{Modelling the rotational light curve}
\label{sec:rota}

\begin{figure}[t]
\resizebox{13.cm}{!}{\includegraphics*{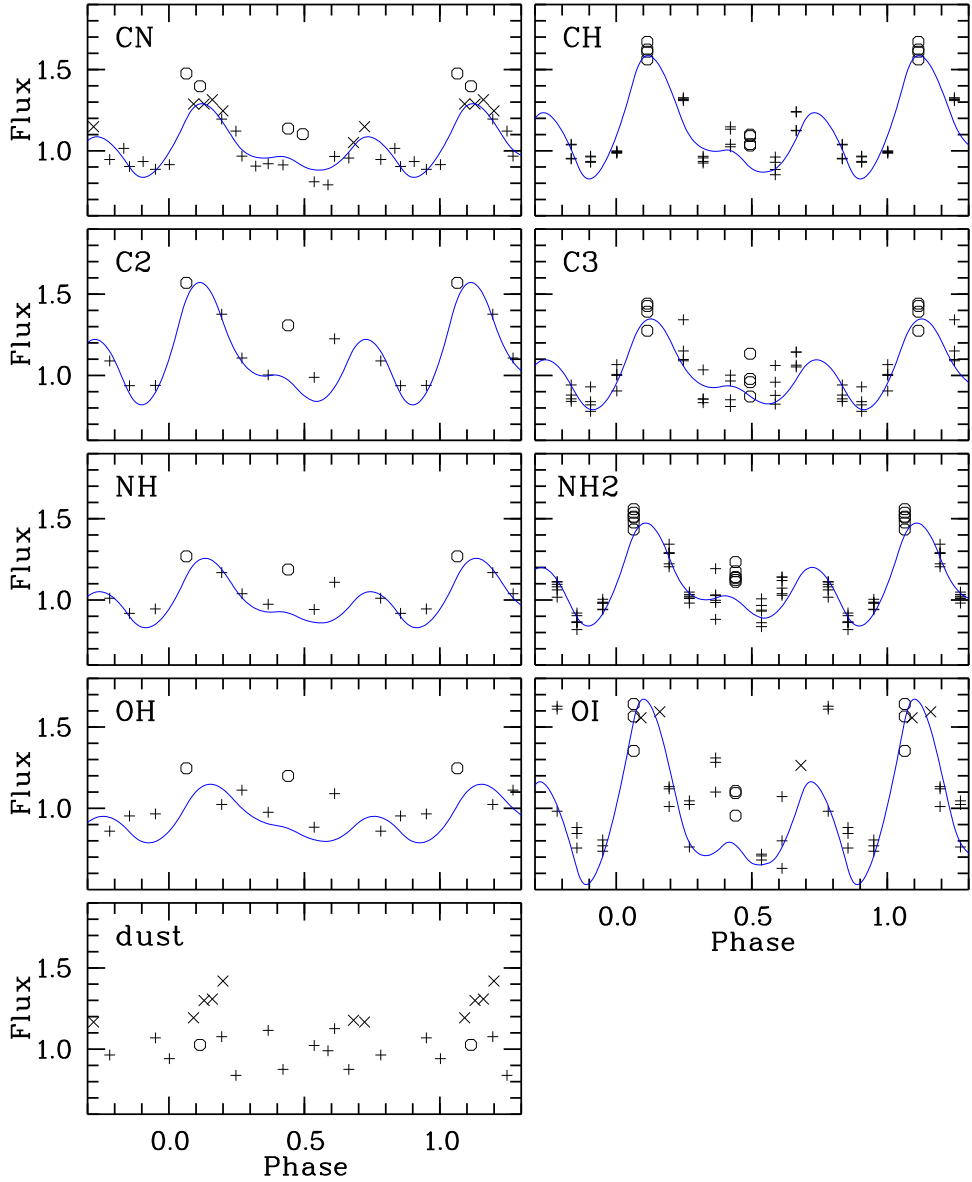}}
      \caption{Phase diagrams of the various species. 
      The phase origin coincides with the impact, and the period
      is 1.709~d. The 
      measurements obtained during the first post-impact
      nights (first night at phase $\sim0.45$, 
      second night at $\sim0.05$) are denoted by circles.
      The curve is a   fit with a three-source model
      optimized for the whole data set, not
      for a particular molecule (see text). 
      Hence some fits may look rather poor, because
      the S/N of the data is bad (e.g., OH).
      The dust does not show
      periodic variations, but the June data appear to be
      systematically higher. The latter are denoted by 
      a different symbol ($\times$). The flux has been 
      scaled to reach $\sim1$ at phase origin.}
  \label{figphase}      
   \end{figure}

A preliminary analysis of the CN light curve from the UVES data
(Fig.~\ref{figphase}) revealed a periodicity of $1.709\pm0.009$~d and
the presence of three main sources or active regions (Jehin et
al. 2006).

We constructed phase diagrams with the above period using the time 
origin of the impact as seen from Earth, July 4, UT 05:52.
In order to allow comparisons, all data were normalized to unity at
impact time, i.e., at phase 0. The spectrum obtained at impact is
strongly affected by twilight and yields noisier data. Hence, the May
data were included in the phase diagram to fix the zero point.

We want to test the hypothesis that the light curve $f_s(t)$ of the
comet in the wavelength of different species ($s$) can be represented
as the response of the comet to a continuous succession of
infinitesimal impact-like events. This is merely a way of describing
the instantaneous production rates of the active regions under the
influence of the solar radiation modulated by the rotation of the
nucleus.  We assume a linear behavior, i.e., $f_s(t)$ is the linear
superposition of the transitory responses to the series of
micro-events:

\begin{equation}
f_s(t) = \int_{-\infty}^t g_s(t') \; n_s(t-t') dt' 
\end{equation}

where $n_s(t)$ is the impact light curve (Eq.~8) corresponding to
species $s$ and $g_s(t)$ represents the total instantaneous activity
of all sources.  $ g_s(t)= g_s(t+P)$ and $f_s(t)=f_s(t+P)$ are
periodic functions of period $P$.  We assume that there is a finite
number $N$ of sources. Their contribution can be described as

\begin{equation}
g_s(t')=\sum_{i=1}^N b_{s,i}(t'-t_{s,i}) + g_{s,0}
\end{equation}
where the $b_{s,i}(t) = b_{s,i}(t+P)$ are the rates of activity of the
$N$ sources, and $g_{s,0}$ is a possible background of activity.

The decomposition into discrete sources arises from physical
considerations, i.e., the idea that the cometary material comes from
several areas or vents. It also allows us to use simple functions $b(t)$
since one might expect the activity of each source to exhibit periodic
rise and fall cycles. 
One may also expect the various
sources to have similar behavior, so the same function could be used
with some phase shift.


It must be emphasized that, because impact and rotational light 
curves overlap for several days, they must be disentangled through an
iterative process, which adds some complexity to the modelling procedure.

The following conclusions were reached using the simplest satisfactory
model, i.e., the one with fewest free parameters.

\begin{itemize}
\item
The two main maxima of the light curves suggest
the presence of at least two sources.  
In fact, three sources are necessary to explain the general shape of the
light curves, i.e., $N=3$. Using only two sources leads to a
pronounced gap in the light curve. 

\item
There is no residual activity  $g_{s,0}=0$

\item
The activity of each source follows the same pattern during the
rotation of the nucleus, i.e., we can write~:
\begin{equation}
 b_{s,i}(t)=a_{s,i} b_s(t)
\end{equation}
with the $b_s(t)$ independent of $i$.

\item
This activity of each source can be approximated by an intermittent
function, a  variant of a
triangular wave.  With $0\leq \tau_{s,1}\leq \tau_{s,2} \leq
P$, we write in the interval $0\leq t\leq P$ :
\begin{equation}
b_s(t)= \left\{ \begin{array}{ll}
 =t/\tau_{s,1}  & \mbox{if $0<t\leq\tau_{s,1}$}\\
 =(\tau_{s,2}-t)/(\tau_{s,2}-\tau_{s,1})  & \mbox{if $\tau_{s,1}<t\leq\tau_{s,2}$}\\
 =0                           & \mbox{if $\tau_{s,2}<t\leq P     $}
  \end{array}
\right.
\end{equation}
This function repeats itself at each interval $k P\leq t\leq (k+1) P$.

\item
The onset times of each source are the same for each species, $t_{s,i}=t_i=(0.18 P,0.47 P,
0.85 P)+kP$
 
\item
The relative intensity of each source is the same for all species,
i.e., $a_{s,i}=c_s d_i$ with $d_i=(0.4,0.7,1.0)$. This is equivalent with
the relative production rates of each species being the same
for each source, i.e., the material is the same. 

\item The long-term characteristic time-scale of the 
decay of $n_s(t)$ is strongly correlated with the 
overall amplitude of the light curve $s$.
This provides an independent  way of predicting  the residuals 
in the post-impact nights.

\item
Finally, $\tau_{s,1}$ and $\tau_{s,2}$, describing the activity of the sources,
are very similar for all
species.  We found that the best fits are obtained with two sets of
parameters.  For CN, CH and [O\,I] we find $\tau_{s,1}=0.21 P$ and
$\tau_{s,2}=0.39 P$ while for OH, NH, NH$_2$, C$_2$, C$_3$,
$\tau_{s,1}=0.19 P$ and $\tau_{s,2}=0.35 P$. The rise and fall
are quasi symmetrical.

\end{itemize}

The above conclusions apply to the gas only. The dust cannot be 
described by the same model.  The phase diagram
shows almost no hint of periodicity whereas a fit with the above
parameters and the response $n_{\mbox{\small dust}}(t)$ observed after the
impact would give very strong variations (see Fig.~\ref{figphase}, bottom).  
This means that either
(i) the temporal variation of the dust luminosity observed after the
impact,
$n_{\mbox{\small dust}}(t)$, is not representative 
of the response of the dust ejected from the sources,
or (ii)
the modulation is masked by a very
large residual component ($g_{\mbox{\small dust},0}$) of non-volatile
dust.

\section{Discussion and conclusions}
\label{sec:conclu}

The high-resolution spectra obtained with the UVES and HIRES spectrographs
have provided photometric data of unexpectedly good quality, in view of the
faintness of the target, the limited area of the slits and the less than perfect
centering. This allowed us to model the impact as well as the rotational light 
curve of this comet which led to the following conclusions.

\begin{itemize}

\item 
Within the uncertainties a simple model reproduces fairly well the
impact light curves of all species. The light curve may reflect
the combination of the production of the observed species and
the expansion of the material throughout the slit.

\item  
All impact light curves are consistent with $\bar{v} \simeq$ 400--600
m/s ($\sigma_v$ = 100--200~m/s).  NH and NH$_2$ have
similar light curves and slightly higher velocities, possibly
indicating a common origin.

\item 
With these expansion velocities, a three-step dissociation sequence
 ``Grand-Parent $\rightarrow$ Parent $\rightarrow$ Daughter'' is
 necessary and sufficient to model all light curves.

\item 
The exact nature of the grandparents is unknown. It can be anything
which desintegrates (or sublimates) as $e^{-t/\tau_g}$. The order 
of magnitude of its
destruction lifetime is typically 10$^{4}$s. It could be common to all
species (e.g., icy dust grains released by the impact).

\item 
Parent lifetimes can only be determined to better than an order of magnitude. 
They largely encompass the various parent lifetimes
proposed in the literature.

\item
The gas activity is generated by three sources, whose duration is less
than about 0.4 of the period, i.e., compatible with a daylight excitation.

\item
The activity of each source follows a similar pattern.

\item 
The progressive increase and decrease of the activity of each source is
compatible with the modulation of the incidence angle of the solar radiation.

\item
The dust observed during the impact disappears rapidly and is probably
mostly sublimating water ice in fragments with a diverse size distribution.

\item 
The dust observed after the impact is dominated by another,
less volatile component, which may consist of the core of
slow-moving dust particles that have lost their icy component and are 
weakly bound to the nucleus.

\item
The material releasing CN, CH and OI seems to be present during a
slightly longer period than that responsible of OH, NH, NH$_2$,
C$_2$, C$_3$.

\end{itemize}

\section{Acknowledgments}
Based on observations carried out at the European Southern Observatory (ESO) 
under prog. 075.C-0355(A), this program was part of a joint European initiative in 
support of the NASA DEEP IMPACT mission to comet 9P/Tempel\,1 by ground-based 
observations at La Silla and VLT Paranal. 
Some of the material presented here is based on data obtained at the W. M. Keck 
Observatory, which immediately made its data available to the public after
impact occurred on July 4, 2005. The W. M. Keck Observatory is operated as a
scientific partnership among the California Institute of Technology, the University
of California and the National Aeronautics and Space Administration.
The Observatory was made possible by the generous financial support of the W.M. 
Keck Foundation. The authors wish to recognize and acknowledge the very significant
cultural role and reverence that the summit of Mauna Kea has always had within
the indigenous Hawaiian community. We are most fortunate to have the opportunity to
get observations from this mountain.
JM is Research Director and DH is Research Associate at FNRS (Belgium).

{}

\clearpage

\renewcommand{\baselinestretch}{1.00}

\begin{table}[h]
\caption{UVES data observational circumstances}
\label{tab:circonsa}
\vspace{.5cm}
\begin{tabular}{lcccrcccccc} \hline\hline 
UT Date (start) & Setup & Exp. & Airm. & Slit PA & Off. & $r$  & $\Delta$ & $\dot{r}$ & $\dot{\Delta}$  & PsAng \\
2005-mm-dd          & (nm)            & (s)         &          & (deg.)  & ('')   &(AU)  & (AU)     & (km/s)    &  (km/s)         & (deg.) \\ \hline\\[-0.2cm]
06-02 / 00:23:14 & 346+580 &  5400 & 1.14 & 117.7 & 0.0 & 1.54 & 0.76 & -3.7 & 5.3 & 117.4 \\ 
06-02 / 01:59:22 & 437+860 &  5400 & 1.17 & 117.7 & 0.0 & 1.54 & 0.76 & -3.6 & 5.5 & 117.4 \\ 
06-07 / 00:36:14 & 346+580 &  5400 & 1.11 & 116.3 & 0.0 & 1.53 & 0.78 & -3.2 & 6.0 & 116.3 \\ 
06-07 / 02:10:49 & 437+860 &  5400 & 1.20 & 0.0   & 0.0 & 1.53 & 0.78 & -3.2 & 6.1 & 116.3 \\ 
06-08 / 00:47:19 & 346+580 &  5400 & 1.11 & 116.0 & 0.0 & 1.53 & 0.78 & -3.1 & 6.2 & 116.0 \\ 
06-08 / 02:27:26 & 437+860 &  5400 & 1.25 & 0.0   & 0.0 & 1.53 & 0.78 & -3.1 & 6.3 & 116.0 \\ 
07-02 / 22:54:56 & 350+580 &  7200 & 1.05 & 21.5  & 1.1 & 1.51 & 0.89 & -0.3 & 8.9 & 115.5 \\ 
07-03 / 00:58:58 & 437+860 &  7200 & 1.11 & 21.5  & 1.2 & 1.51 & 0.89 & -0.3 & 9.2 & 115.5 \\ 
07-03 / 22:54:01 & 350+580 &  7200 & 1.05 & 21.5  & 1.0 & 1.51 & 0.89 & -0.2 & 9.0 & 111.4 \\ 
07-04 / 00:58:31 & 437+860 &  7200 & 1.11 & 21.5  & 1.4 & 1.51 & 0.89 & -0.1 & 9.0 & 111.4 \\ 
07-04 / 22:54:47 & 348+580 &  7200 & 1.05 & 21.5  & 0.9 & 1.51 & 0.90 & 0.0 & 9.2 &  111.3 \\ 
07-05 / 01:01:27 & 437+860 &  7800 & 1.12 & 111.5 & 1.9 & 1.51 & 0.90 & 0.0 & 9.4 &  111.3 \\ 
07-06 / 00:10:42 & 348+580 &  9600 & 1.05 & 135.0 & 1.3 & 1.51 & 0.90 & 0.1 & 9.5 &  111.1 \\ 
07-06 / 02:54:53 & 437+860 &  4800 & 1.60 & 135.0 & 0.0 & 1.51 & 0.90 & 0.1 & 9.7 &  111.1 \\ 
07-06 / 22:55:07 & 348+580 &  7500 & 1.04 & 111.0 & 0.0 & 1.51 & 0.91 & 0.2 & 9.4 &  111.0 \\ 
07-07 / 01:03:45 & 437+860 &  7500 & 1.13 & 111.0 & 0.0 & 1.51 & 0.91 & 0.2 & 9.6 &  111.0 \\ 
07-07 / 22:50:59 & 348+580 &  7500 & 1.04 & 21.0  & 0.0 & 1.51 & 0.91 & 0.3 & 9.6 &  110.9 \\ 
07-08 / 00:59:17 & 437+860 &  7500 & 1.12 & 21.0  & 0.0 & 1.51 & 0.91 & 0.3 & 9.8 &  110.9 \\ 
07-08 / 22:54:05 & 348+580 &  7500 & 1.03 & 21.0  & 0.0 & 1.51 & 0.92 & 0.4 & 9.7 &  110.7 \\ 
07-09 / 01:03:09 & 437+860 &  7500 & 1.13 & 21.0  & 0.0 & 1.51 & 0.92 & 0.4 & 9.9 &  110.7 \\ 
07-09 / 22:51:16 & 348+580 &  7800 & 1.03 & 21.0  & 0.0 & 1.51 & 0.92 & 0.5 & 9.8 &  110.6 \\ 
07-10 / 01:05:21 & 437+860 &  7800 & 1.14 & 21.0  & 0.0 & 1.51 & 0.93 & 0.6 & 10.0&  110.6 \\ 
07-10 / 22:52:20 & 348+580 &  7200 & 1.03 & 21.0  & 0.0 & 1.51 & 0.93 & 0.7 & 9.9 &  110.5 \\ 
07-11 / 00:59:27 & 437+860 &  7200 & 1.12 & 21.0  & 0.0 & 1.51 & 0.93 & 0.7 & 10.2 & 110.5 \\ 
07-11 / 22:54:57 & 348+580 &  7200 & 1.03 & 21.0  & 0.0 & 1.51 & 0.94 & 0.8 & 10.0 & 110.3 \\ 
07-12 / 00:58:51 & 437+860 &  7200 & 1.13 & 21.0  & 0.0 & 1.51 & 0.94 & 0.8 & 10.3 & 110.3 \\ 
\hline
\end{tabular}
\vspace{2mm}

{\small $r$ is the heliocentric distance in astronomical units (AU) and $\Delta$ is the 
geocentric distance, $\dot{r}$ is the heliocentric radial velocity, Exp. is the 
the total exposure time in seconds, Airm is the air-mass at mid exposure, Slit PA is
the slit position angle (North to East), Off. is the offset of the slit center from the 
nucleus, PsAng is the position angle of the Sun-Comet vector. The Setup column 
defines the central wavelength of the UVES blue and red arm.}
\end{table}

\begin{table}[h]
\caption{HIRES data observational circumstances}
\label{tab:circonsb}
\vspace{.5cm}
\begin{tabular}{lrccccccc} \hline\hline 
UT Date (start) & Exp.  & Airm. & Slit PA & $r$  & $\Delta$ & $\dot{r}$ & $\dot{\Delta}$  &  PsAng  \\
2005-mm-dd      & (s)         &          &  (deg.) & (AU)    & (AU)     & (km/s)    &  (km/s) & (deg.) \\ \hline\\[-0.2cm]
05-30 / 08:33:06 &	  1200 & 1.19 & 59.3 & 1.55 & 0.75 & -4.0 & 5.1 &  118.6   \\
05-30 / 08:54:02 &	  1200 & 1.26 & 63.3 & 1.55 & 0.75 & -4.0 & 5.1 &  118.6   \\
05-30 / 09:14:55 & 	  1200 & 1.35 & 66.3 & 1.55 & 0.75 & -4.0 & 5.2 &  118.6   \\
07-04 / 05:36:15 &	   720 & 1.16 & 11.6 & 1.51 & 0.89 & -0.1 & 9.1 &  111.4   \\
07-04 / 05:55:18 &	   600 & 1.18 & 20.1 & 1.51 & 0.89 & -0.1 & 9.1 &  111.4   \\
07-04 / 06:06:12 &	   600 & 1.19 & 24.8 & 1.51 & 0.89 & -0.1 & 9.2 &  111.4   \\
07-04 / 06:17:05 & 	   900 & 1.22 & 29.7 & 1.51 & 0.89 & -0.1 & 9.2 &  111.4   \\
07-04 / 06:32:59 &	   900 & 1.25 & 35.5 & 1.51 & 0.89 & -0.1 & 9.2 &  111.4   \\
07-04 / 06:48:52 &	   900 & 1.29 & 40.6 & 1.51 & 0.89 & -0.1 & 9.2 &  111.4   \\
07-04 / 07:04:46 &	   900 & 1.35 & 45.0 & 1.51 & 0.89 & -0.1 & 9.3 &  111.4   \\
07-04 / 07:20:41 &	   900 & 1.41 & 48.9 & 1.51 & 0.89 & -0.1 & 9.3 &  111.4   \\
07-04 / 07:36:35 &	   900 & 1.49 & 52.3 & 1.51 & 0.89 & -0.1 & 9.3 &  111.4   \\
07-04 / 07:52:29 &	   900 & 1.59 & 55.2 & 1.51 & 0.89 & -0.1 & 9.3 &  111.4   \\
07-04 / 08:08:25 &	   900 & 1.71 & 57.8 & 1.51 & 0.89 & -0.1 & 9.4 &  111.4   \\
07-04 / 08:24:19 &	   900 & 1.87 & 60.0 & 1.51 & 0.89 & -0.1 & 9.4 &  111.4   \\
07-04 / 08:40:13 &	  1800 & 2.19 & 65.3 & 1.51 & 0.89 & -0.1 & 9.4 &  111.4   \\
07-04 / 09:11:09 &	  1800 & 2.88 & 68.6 & 1.51 & 0.89 & -0.1 & 9.4 &  111.3   \\
\hline
\end{tabular}
\end{table}

\begin{table}[h]
\caption{{The gaseous species analyzed} 
 }
\label{tab:gasspec}
\vspace{.5cm}
\begin{tabular}{lccc} \hline\hline
Species & Central  windows & Side windows      & Extinction  \\
        &                  & Blue side ~ ~ ~ Red side &   Paranal / Mauna Kea  \\ 
         &  (\AA)         & (\AA)               & (mag/airm)  \\ \hline\\[-0.2cm]
OH       & 3093.4-  96.8 & 3092.0-  93.2 ~  3097.0-  98.0 &   1.60 / 1.37 \\
NH       & 3353.2-  54.4 & 3352.2-  53.2 ~  3354.2-  55.0 &   0.62 / 0.54 \\ 
         & 3357.2-  58.8 & 3355.5-  57.2 ~  3358.8-  60.0 &   0.62 / 0.54 \\
         & 3364.6-  65.5 & 3363.4-  64.5 ~  3365.3-  66.3 &   0.62 / 0.54 \\ 
         & 3368.8-  69.5 & 3367.5-  68.5 ~  3369.5-  70.4 &   0.62 / 0.54 \\
CN       & 3876.1-  82.0 & 3874.9-  75.7 ~  3882.7-  83.7 &   0.35 / 0.29 \\
C$_3$    & 4048.9-  55.0 & 4047.4-  48.9 ~  4054.4-  56.9 &   0.27 / 0.24  \\
         & 4051.3-  52.2 & 4047.9-  49.4 ~  4053.4-  54.9 &   0.27 / 0.24  \\
         & 4071.9-  75.9 & 4070.9-  71.9 ~  4075.9-  76.9 &   0.27 / 0.24  \\
CH       & 4299.9-  00.6 & 4298.9-  99.9 ~  4299.6-  01.2 &   0.22 / 0.19  \\
         & 4303.9-  04.1 & 4301.9-  02.9 ~  4304.4-  05.4 &   0.22 / 0.19  \\
C$_2$    & 5164.5-  65.4 & 5163.2-  64.4 ~  5165.4-  66.8 &   0.14 / 0.13 \\
NH$_2$   & 5185.9-  86.4 & 5184.8-  85.8 ~  5186.8-  87.8 &   0.14 / 0.13 \\
         & 5193.8-  94.3 & 5192.8-  93.8 ~  5194.3-  95.3 &   0.14 / 0.13 \\
         & 5731.4-  32.1 & 5729.8-  30.6 ~  5732.3-  33.3 &   0.12 / 0.12 \\
$[$O\,I$]$ & 5577.0-  77.5 & 5576.0-  77.0 ~  5578.0-  79.0 &   0.13 / 0.12 \\
         & 6299.5-  01.0 & 6299.0-  99.5 ~  6301.0-  02.0 &   0.07 / $-$ ~ \\
\hline
\end{tabular}
\end{table}

\begin{table}[h]
\caption{Results of model fitting}
\label{tab:results}
\begin{center}
\begin{tabular}{lcccc}\hline\hline 
       & $\bar{v}$ &  $\sigma_v$   &  $\tau_p$       &  $\tau_p / \tau_g$\\ 
       & (m/s)     &  (m/s)        &  (s)            &   \\ \hline\\[-0.2cm]
CN     & 400       & 100--200      & 2--50   10$^{4}$  & 1.1--100 \\
OH     & 400       & 200           & 2--40   10$^{4}$  & 1.1--100   \\
NH$_2$ & 500       & 100           & 6--30   10$^{3}$  & 0.01--0.9  \\
C$_3$  & 400       & 200           & 8--30   10$^{3}$  & 0.01--0.9  \\
CH     & 400       & 200           & 1--20   10$^{4}$  & 1.1--100   \\
NH     & 550       & 150           & 3--100  10$^{4}$  & 1.1--100   \\
C$_2$  & 650       & 200           & 4--200  10$^{4}$  & 1.1--100    \\
OI     & 400       & 100           & 6--90   10$^{3}$  & 1.1--50    \\
 \hline
\end{tabular}
\end{center}
\end{table}

\clearpage

\renewcommand{\baselinestretch}{2.00}

\clearpage

\end{document}